\documentclass[twocolumn,showpacs,preprintnumbers,amsfonts,amsmath,amssymb,nofootinbib,floatfix]{revtex4-1}

\usepackage{graphicx}
\usepackage{dcolumn}
\usepackage{bm}

\renewcommand{\bf}{\boldsymbol}
\newcommand{\be}{\begin{equation}}
\newcommand{\ee}{\end{equation}}
\newcommand{\ba}{\begin{eqnarray}}
\newcommand{\ea}{\end{eqnarray}}
\newcommand{\bi}{\begin{itemize}}
\newcommand{\ei}{\end{itemize}}

\newcommand{\Tr}{\,\hbox{\rm Tr}}

\newcommand{\msbar}{{\overline{\mbox{\scriptsize MS}}}}

\begin{document}

\title{Equation of state of a relativistic theory from a moving frame}

\author{Leonardo Giusti$^{a,b}$, Michele Pepe$^b$}

\affiliation{\vspace{0.3cm}
\vspace{0.1cm} 
$^a$ Dipartimento di Fisica, Universit\`a di Milano-Bicocca, Piazza della Scienza 3, 
I-20126 Milano, Italy\\
$^b$ INFN, Sezione di Milano-Bicocca, Piazza della Scienza 3, 
I-20126 Milano, Italy}


\begin{abstract}
We propose a new strategy for determining the equation of state of a relativistic 
thermal quantum field theory by considering it in a moving reference system. In this 
frame an observer can measure the entropy density of the system directly 
from its average total momentum. In the Euclidean path integral formalism, this amounts 
to compute the expectation value of the off-diagonal components $T_{0k}$ of the 
energy-momentum tensor in presence of shifted boundary conditions. The entropy is thus 
easily measured from the expectation value of a local observable computed at the target temperature 
$T$ only. At large $T$, the temperature itself is the only scale which drives the systematic 
errors, and the lattice spacing can be tuned to perform a reliable continuum limit 
extrapolation 
while keeping finite-size effects under control. We test this strategy for the four-dimensional 
$SU(3)$ Yang-Mills theory. We present precise results for the entropy density and its 
step-scaling function in the temperature range $0.9 \,T_c - 20\, T_c$. At each temperature, 
we consider four lattice spacings in order to extrapolate the results to the continuum limit. As a byproduct 
we also determine the ultraviolet finite renormalization constant of $T_{0k}$ by imposing suitable 
Ward identities. These findings establish this strategy as a solid, simple and efficient method 
for an accurate determination of the equation of state of a relativistic thermal field theory 
over several orders of magnitude in $T$.
\end{abstract}

\maketitle


\noindent {\it Introduction.---} 
Relativistic thermal quantum field theories are of central importance in 
many areas of research in physics. The equation of state (EOS) of Quantum 
Chromo Dynamics (QCD) is a very basic property of strongly-interacting 
matter that is of absolute interest in particle and nuclear physics, and 
in cosmology. It is also a crucial input in the analysis of data collected 
at the heavy-ion colliders.\\
\indent Lattice QCD is the only known theoretical framework where the EOS can be determined 
from first principles in the interesting range of temperature values. Since 
the perturbative expansion converges very slowly, the full computation of the 
EOS has to be done numerically over several orders of magnitude in $T$. 
Severe unphysical contributions hinder the standard way of computing the pressure and 
the energy density. The expansion of the free energy in the bare parameters, and the 
subtraction of ultraviolet power divergences make the computation of the EOS technically 
difficult and numerically very demanding~\cite{Boyd:1996bx,Borsanyi:2012ve,Cheng:2009zi,Borsanyi:2013bia} (see
Ref.~\cite{Philipsen:2012nu} for a recent review). Temperatures higher than a few hundreds
MeV are still unreachable with staggered fermions. The computation remains prohibitive
with Wilson fermions. The obstacles, however, are not rooted in the physics content of the EOS, 
but in the strategy adopted for its computation.  This calls for a conceptual progress 
able to trigger new computational strategies, which in turn are capable to reach the goal 
of a precise computation of the EOS in a generic discretization of the theory.\\
\indent The underlying Lorentz symmetry of relativistic thermal theories 
offers an elegant and simple solution to this problem. In these theories the 
entropy is proportional to the total momentum of the system as measured by
an observer in a moving frame. Remarkably, the corresponding Euclidean path integral formulation
is rather simple. It corresponds to inserting a shift ${\bf \xi}$ in the spatial directions when
closing the boundary conditions of a field $\phi$ in the compact direction of length
$L_0$~\cite{Giusti:2012yj,Giusti:2011kt,Giusti:2010bb,DellaMorte:2010yp}
\vspace{-0.625cm}

\begin{equation}
\phi (L_0,{\bf x}) = \phi (0,{\bf x} - L_0\, {\bf \xi})\; .
\end{equation}
\vspace{-0.625cm}

\noindent
In the thermodynamic limit, the invariance of the dynamics under
the $SO(4)$ group implies that the free energy density $f(L_0 ,{\bf \xi})$
satisfies~\cite{Giusti:2012yj,Giusti:2011kt,Giusti:2010bb}
\vspace{-0.5cm}
 
\begin{equation}\label{eq:freeE}
f(L_0 ,{\bf \xi}) = f(L_0 \sqrt{1+{\bf \xi}^2},{\bf 0})\; .
\end{equation}
\vspace{-0.5cm}

\noindent
Hence the free energy does not depend on $L_0$ and ${\bf \xi}$ separately but on the
combination $L_0 \sqrt{1+{\bf \xi}^2}= T^{-1}$ which fixes the inverse temperature of the system.
This redundancy implies that the thermal distributions of the total energy and momentum are
related, and interesting Ward identities (WIs) follow.
In particular, the entropy density can be written 
as~\cite{Giusti:2012yj}
\vspace{-0.625cm}

\begin{equation}\label{eq:sxi}
\frac{s(T)}{T^3} =  -\frac{(1+ {\bf \xi}^2)}{\xi_k} 
\frac{\langle T_{0k} \rangle _{\bf \xi}}{T^4}\; ,
\end{equation}
\vspace{-0.5cm}

\noindent where $\langle \cdot \rangle _{\bf \xi}$ stands for the expectation 
value computed with a non-zero shift ${\bf \xi}$. No
ultraviolet power-divergent contributions need to be subtracted
from $\langle T_{0k} \rangle _{\bf \xi}$.

In this Letter we explore a new computational strategy for determining the EOS 
of a relativistic thermal quantum field theory based on Eq.~(\ref{eq:sxi}). 
We illustrate the power of the method in the $SU(3)$ Yang-Mills theory, where 
we determine the entropy density of the system in the range $0.9\, T_c - 20\,T_c$. 
This is a particularly interesting theory since it is the limit of QCD in absence 
of fermions (or with infinitely heavy fermions), and it can be used to test new 
ideas and numerical methods without facing the problems of simulating 
dynamical fermions. Since it relies on Lorentz invariance only, the strategy is 
directly applicable to any relativistic thermal theory and, in particular, to QCD.\\
\noindent {\it Entropy density from the lattice.---} 
We regularize the four-dimensional $SU(3)$ Yang--Mills theory on a square
lattice of size $ L_0 \times L^3$ and of spacing $a$. The link variables 
$U_\mu(x) \in SU(3)$ represent the gauge field and the Wilson action $S$ 
is, up to a constant, given by
\vspace{-0.5cm}

\begin{equation}
S[U] = -\frac{\beta}{6}\! \sum_{x,\mu\nu} 
\!\mbox{Re} \Tr [U_\mu(x)U_\nu(x+\hat\mu)U^\dagger_\mu(x+\hat\nu) 
U^\dagger_\nu(x)]\nonumber
\end{equation}
\vspace{-0.5cm}

\noindent where $\beta=6/g_0^2$, and $g_0$ is the bare coupling.
We impose periodic boundary conditions in the spatial directions and shifted boundary 
conditions along the compact direction, 
$U_\mu(L_0,{\bf x}) = U_\mu(0,{\bf x}- L_0\, {\bf \xi})$,
where $(L_0/a)\, {\bf \xi}$ is a vector with integer components. We consider the clover definition of 
the energy-momentum tensor on the lattice~\cite{Caracciolo:1989pt}
\vspace{-0.5cm}

\be\label{eq:TmunuLat}
T_{\mu\nu} =  \frac{\beta}{6}\Big\{F^a_{\mu\alpha}F^a_{\nu\alpha}
- \frac{1}{4} \delta_{\mu\nu} F^a_{\alpha\beta}F^a_{\alpha\beta} \Big\}\; .
\ee
\vspace{-0.5cm}

\noindent The field strength tensor is defined as 
\vspace{-0.5cm}

\be
F^a_{\mu\nu}(x) = - \frac{i}{4 a^2} 
\Tr\Big\{\Big[Q_{\mu\nu}(x) - Q_{\nu\mu}(x)\Big]T^a\Big\}\; ,  
\ee
\vspace{-0.5cm}

\noindent where $T^a=\lambda^a/2$ with $\lambda^a$ being the Gell-Mann matrices, and 
(see Ref.~\cite{Caracciolo:1989pt} for more details)
\vspace{-0.5cm}

\begin{equation}
Q_{\mu\nu}(x) = P_{\mu\nu}(x) + P_{\nu-\mu}(x) + P_{-\mu-\nu}(x) +P_{-\nu\mu}(x)\; .
\end{equation}
\vspace{-0.5cm}

\noindent The matrix $P_{\mu\nu}(x)$ is the parallel transport along an elementary plaquette at the
lattice site $x$ along the directions $\mu$ and $\nu$, and the minus sign stands for the
negative orientation. The lattice regularization breaks explicitly translation 
invariance down to a discrete sub-group. As a consequence the off-diagonal components of the 
energy-momentum tensor renormalize multiplicatively~\cite{Caracciolo:1989pt}, and 
Eq.~(\ref{eq:sxi}) becomes
\vspace{-0.5cm}

\be\label{eq:Snorm}
\frac{s(T)}{T^3} =  -\frac{(1+{\bf \xi}^2)}{\xi_k} \,
 \frac{Z_T \langle T_{0k} \rangle _{\bf \xi}}{T^4}\; .
\ee
\vspace{-0.25cm}

\begin{figure*}[!t]
\vspace{-1.5cm}

\begin{tabular}{cc}
\includegraphics[width=8.0 cm,angle=0]{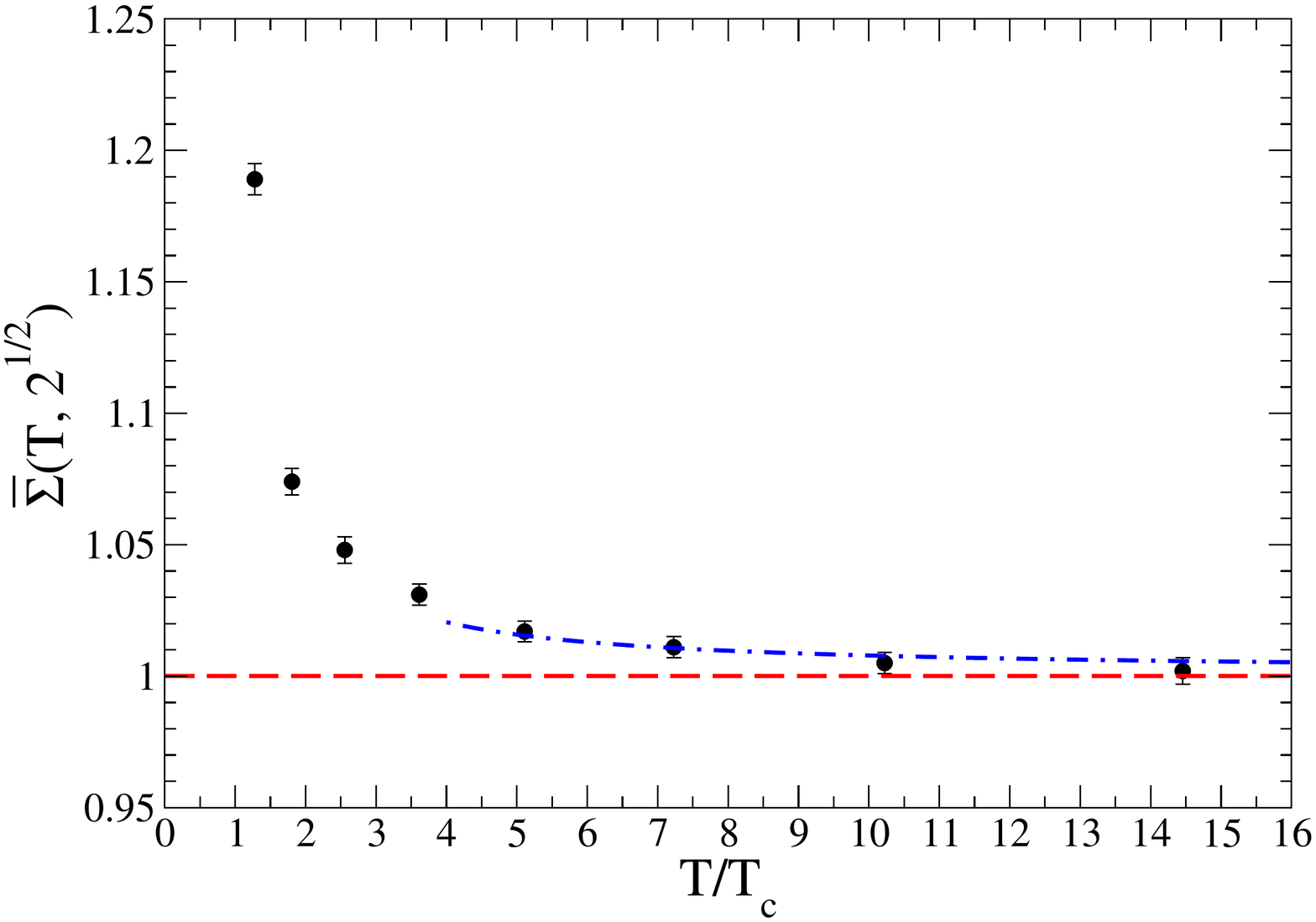} &
\includegraphics[width=8.0 cm,angle=0]{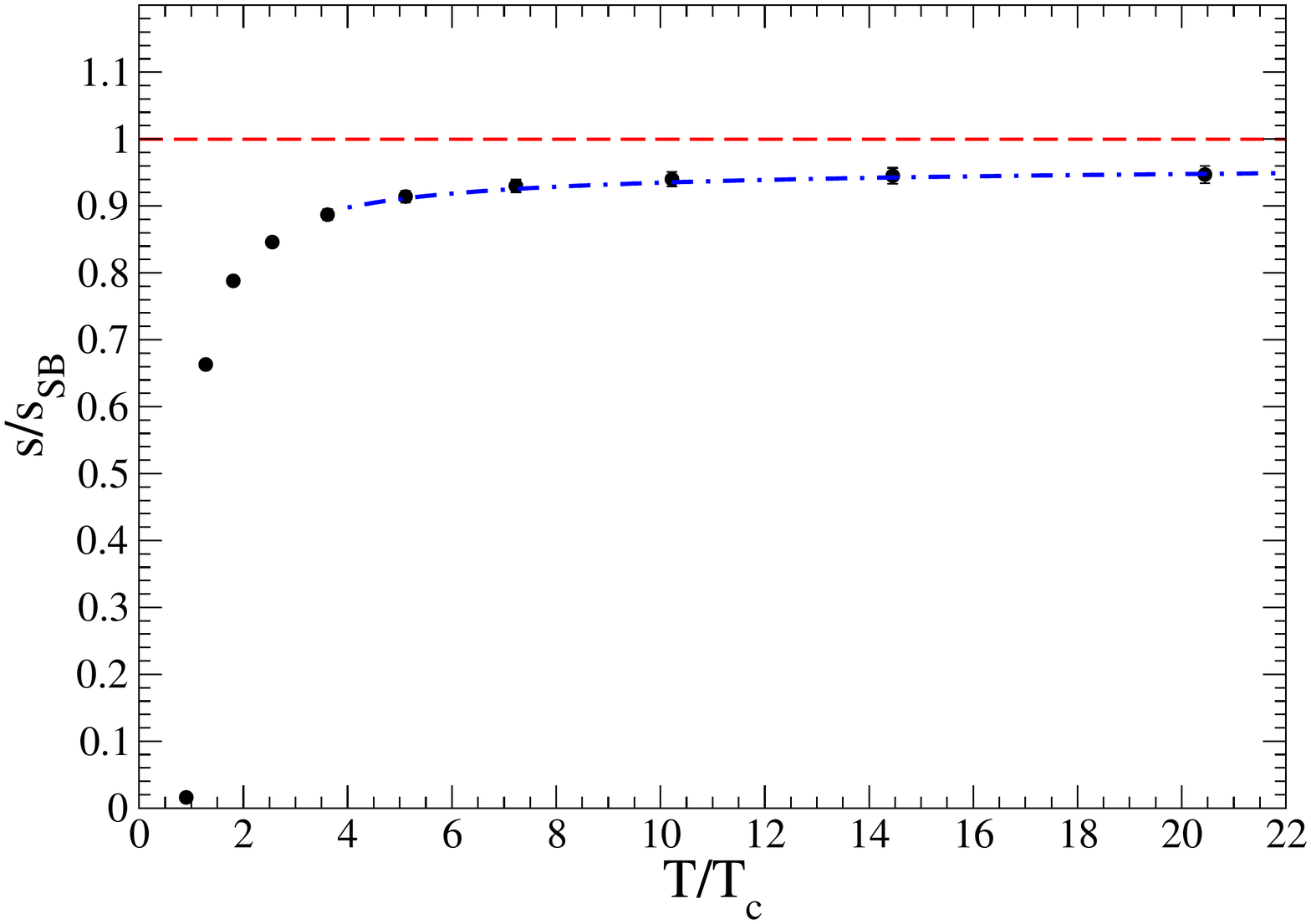}\\
\end{tabular}
\vspace{-0.625cm}

\caption{The step-scaling function (left) and the entropy density normalized to the 
SB value (right) versus 
the temperature. The dashed lines (red) are the SB values, while the dotted-dashed lines (blue)
are the perturbation theory ones from Ref.~\cite{Kajantie:2002wa}.\label{fig:cont_res}}
\vspace{-0.5cm}

\end{figure*}

\noindent The renormalization constant $Z_T$ of $T_{0k}$ can be fixed by imposing suitable 
WIs \cite{Giusti:2011kt,Giusti:2012yj}. $Z_T$ depends only on the bare coupling constant 
and, up to discretization effects, it is independent of the 
kinematic parameters e.g., $L$, $T$, ${\bf \xi}$. 
These parameters can be chosen at will, with the condition that they remain constant in physical 
units when approaching the continuum limit, or that they generate in $Z_T$
negligible discretization effects compared to the statistical errors. 
Ultimately, which WI and/or kinematics are the most effective has to be investigated numerically. 
We have found that for the $SU(3)$ Yang--Mills theory discretized with the Wilson action, 
$Z_T$ can be determined with small discretization
effects and with a limited numerical effort as
\vspace{-0.5cm}

\be\label{eq:Znp}
Z_T = \frac{1}{2 a L^{3}}
\frac{1}{\langle T_{0k} \rangle_{{\bf \xi}}}
\ln{\frac{Z(L_0,{\bf \xi} + a/L_0 \hat k)}{Z(L_0,{\bf \xi} - a/L_0 \hat k)}}\; , 
\ee
\vspace{-0.25cm}

\noindent where $Z(L_0,{\bf \xi})$ is the partition function of the theory.
Once $Z_T$ is known, the lattice size and spacing can be adjusted so to
carry out a reliable continuum limit extrapolation of the entropy density at any given
value of $T$ with moderate computational resources. This is possible thanks to the fact
that at large $T$ the temperature itself is the 
only relevant scale that drives discretization and finite volume effects. The
mass gap of the theory is proportional to $T$, and small pre-factors in its
expression do not invalidate the strategy. Indeed increasing the spatial size of the lattice
does not increase the computational effort at fixed statistical accuracy 
since $T_{0k}$ is a local observable.\\
\noindent A slightly different approach is to define a step-scaling function 
$\Sigma (T,r)$ for the entropy density as 
\vspace{-0.5cm}

\begin{equation}\label{eq:stepfun}
\Sigma (T,r) = \frac{ T^3 s(T')}{T'^3 s(T)} = 
\frac{(1+{\bf \xi} ^{' 2})^3\,\, \xi_k}{(1+\bf \xi ^2)^3\,\, \xi'_k} 
\frac{\langle T_{0k} \rangle _{\bf \xi'} }{\langle T_{0k} \rangle _{\bf \xi} }\; , 
\end{equation}
\vspace{-0.25cm}

\noindent where ${\bf \xi}$ and ${\bf \xi}'$ are two different shifts. The factor $Z_T$ drops out
and the step-scaling function has a universal continuum limit as it stands. When $L_0$ and
$\beta$ are kept fixed, the step $r$ in the temperature is given by the ratio $r=T'/T=
\sqrt{1+{\bf \xi}^2}/\sqrt{1+{\bf \xi} ^{' 2}}$.  Once $\Sigma(T,r)$ is known, the entropy
density at a given temperature can be obtained from its value at a single reference 
temperature $T_0$ by solving the straightforward recursion relation. Thus, $Z_T$ has to be 
determined only at the values of $\beta$
where $s(T_0)/T_0^3$ is being measured.\\
\begin{figure*}[thb]
\vspace{-1.5cm}

\begin{center}
\begin{tabular}{cc}
\includegraphics[width=8.0 cm,angle=0]{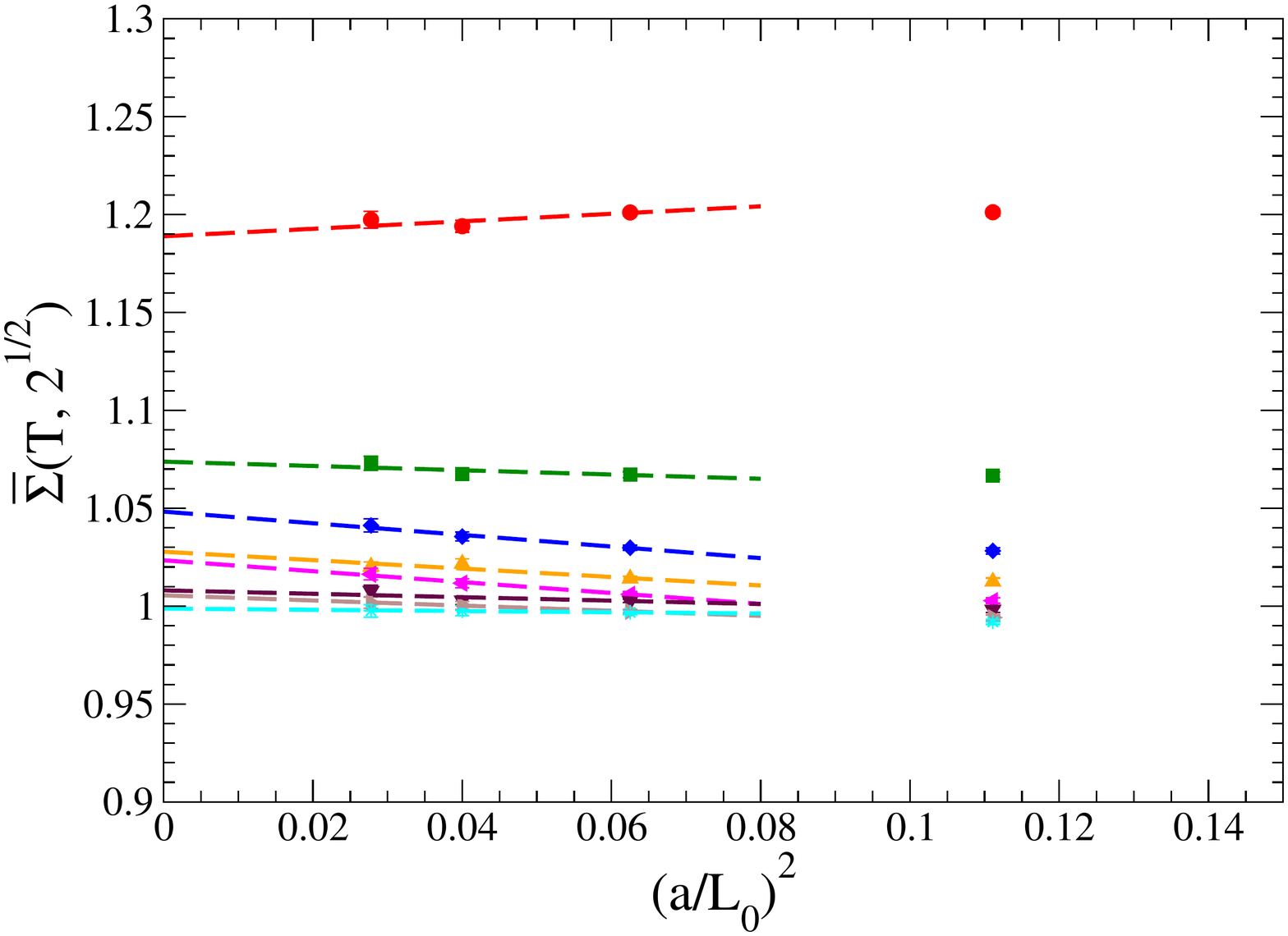} &
\includegraphics[width=8.0 cm,angle=0]{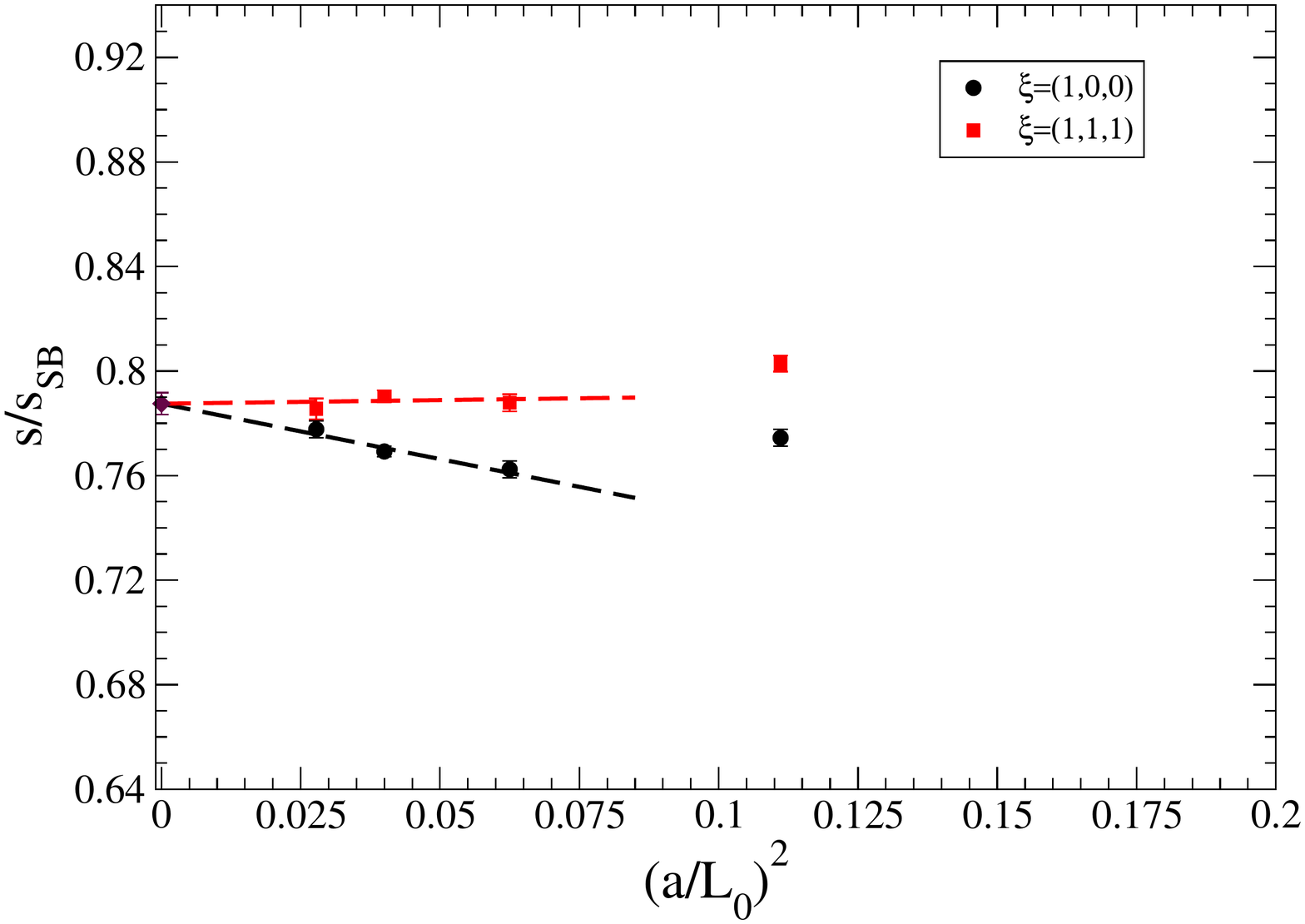}\\
\end{tabular}
\vspace{-0.625cm}

\caption{Left: continuum limit extrapolation of the entropy step-scaling function. Right: 
continuum limit extrapolation of entropy density at the reference temperature $T_0$, 
normalized to the Stefan-Boltzmann (SB) value $s_{\rm SB}/T^3= 32\pi^2/45$.\label{fig:cont_extr}}
\end{center}
\vspace{-0.75cm}

\end{figure*}
\noindent {\it Numerical computation.---} 
We have measured the entropy density (preliminary results were 
presented in \cite{Giusti:2013mxa}) in the 
range $0.9\, T_c - 20\, T_c$, where $T_c$ is the critical temperature. 
We opted for computing the step-scaling function at 9 temperatures in the range 
$T_0/2\,$--$\,8\, T_0$, with values separated by a step-factor of about $\sqrt{2}$. 
The reference temperature has been fixed to $T_0=L^{-1}_{\rm max}$, where $L_{\rm max}$ in 
units of the standard reference scale $r_0$ corresponds to 
$L_{\rm max}/r_0=0.738(16)$~\cite{Capitani:1998mq,Necco:2001xg}. The critical temperature
is $r_0 T_c=0.750(4)$~\cite{Boyd:1996bx,Lucini:2003zr}, and therefore 
$T_0\simeq 1.807\, T_c$. At this temperature we have computed also the renormalization 
constant $Z_T$.
At each value of the lattice spacing and of $L_0/a$, we have measured 
$\langle T_{0k} \rangle _{\bf \xi}$ for two shifts, ${\bf \xi}=(1,0,0)$ and $(1,1,1)$
with standard numerical techniques. The step-scaling function is then computed 
by using Eq.~(\ref{eq:stepfun}) as 
$\Sigma\big(1/(2 L_0),\sqrt{2}\big)= 
\langle T_{0k} \rangle _{(1,0,0)}/(8 \langle T_{0k} \rangle _{(1,1,1)})$ 
At each $T$ we have collected data at four different values of the 
lattice spacing, corresponding to $L_0/a=3$, $4$, $5$ and $6$. At the first four 
temperatures, $\beta$ has been fixed from $r_0/a$ by requiring that 
$L_{\rm max}=0.738\, r_0$~\cite{Necco:2001xg}.  For the other data sets, we have
determined $\beta$ by interpolating quadratically in $\ln{(L/a)}$ the data listed in
Tables A.1 and A.4 of Ref.~\cite{Capitani:1998mq} corresponding to fixed values of 
$\bar g^2(L)$. In order to keep finite volume effects below the statistical errors,
we have considered $T L \geq 12$. Taking into account the present estimate of the lightest
screening mass, finite size effects are expected to be negligible compared to our statistical
errors~\cite{Giusti:2012yj}. On the coarsest lattice of each data set, 
we have performed numerical simulations at a smaller volume. No finite size corrections were 
observed within errors. All the details of the simulations will 
be reported elsewhere~\cite{noi:2014big}. We just note that the $\beta$ values 
range from $5.85$ to $8.6$, and the number of lattice points in the spatial directions 
goes from $64^3$ to $128^3$.\\ 
\begin{table}[t!]
\begin{center}
\begin{tabular}{lcc}
\hline\\[-0.175cm]
$T/T_0$  &$\Sigma_s(T,\sqrt{2})$&$s/s_{\rm SB}$\\[0.125cm]
\hline\\[-0.175cm]
$1/2$       & $42(9)$    & $0.016(3)$\\[0.125cm]
$1/\sqrt{2}$& $1.189(6)$ & $0.663(5)$\\[0.125cm]
$1$         & $1.074(5)$ & $0.788(4)$\\[0.125cm]
$\sqrt{2}$  & $1.048(5)$ & $0.846(6)$\\[0.125cm]
$2$         & $1.031(4)$ & $0.887(8)$\\[0.125cm]
$2\sqrt{2}$  &$1.017(4)$ & $0.914(9)$\\[0.125cm]
$4        $  &$1.011(4)$ & $0.930(10)$\\[0.125cm]
$4\sqrt{2}$  &$1.005(4)$ & $0.940(11)$\\[0.125cm]
$8        $  &$1.002(5)$ & $0.945(12)$\\[0.125cm]
$8\sqrt{2}$  &  -        & $0.947(13)$\\[0.125cm]
\end{tabular}
\vspace{-0.375cm}

\caption{
\label{tab:continuum} Continuum limit extrapolated values of the step-scaling
function and of the entropy density.}
\end{center}
\vspace{-1.0cm}

\end{table}
\noindent In Fig.~\ref{fig:cont_extr} we show the results for 
$\bar\Sigma=\Sigma-\Sigma_0+1$ as a function of $(a/L_0)^2$ for the 8
highest temperatures, where $(\Sigma_0-1)$ are the tree-level discretization 
effects that are subtracted analytically~\cite{Giusti:2012yj}. The statistical errors
range from 1 per-mille up to 3.5 per-mille. For these data 
sets
the residual lattice artifacts turn out to be very small, 
and at most of $2\%$ already at $L_0/a=3$. A continuum linear extrapolation
in $(a/L_0)^2$ of the three points with finer lattice spacings works very well for all data sets as shown 
in Fig.~\ref{fig:cont_extr}. The intercepts of these fits are our best estimate of the 
step-scaling function in the continuum limit. A quadratic fit of all four points give always compatible 
results within the statistical errors. The same applies for a combined fit of all data with discretization 
effects parametrized as expected in the weak coupling expansion. For the last 5 temperatures we interpolate 
the results for $\Sigma_s(T,\sqrt{2})$ in the renormalized coupling, and use the 
fit function to correct for the slight mismatch in the scales from 
Ref.~\cite{Capitani:1998mq}. The best values for the step-scaling function are 
given in Tab.~\ref{tab:continuum}, and shown in the left plot of Fig.~\ref{fig:cont_res}.\\ 
\indent The renormalization constant $Z_T$ has been determined from 
Eq.~(\ref{eq:Znp}). In this case it is not
necessary to consider large spatial volumes, and the numerical simulations have been
performed with $L/a=12$ and $16$.  The finite-volume $\langle T_{0k} \rangle _{\bf \xi}$ in the 
denominator has been computed as described above. The derivative in the numerator 
requires the calculation of a ratio of two partition functions which cannot be computed in 
a single Monte Carlo simulation due to the very poor overlap of the relevant phase space 
of the two integrals. In this case we have used the Monte Carlo procedure of 
Refs.~\cite{Giusti:2010bb,DellaMorte:2010yp}. We consider a set of $(n+1)$ systems with action
$\overline S(U,r_i)= r_i S(U^{({\bf \xi} - a/L_0 \hat k)}) + (1-r_i) S(U^{({\bf \xi} + a/L_0 \hat k)})$
($r_i=i/n$, $i=0,1,\dots,n$), where the superscript indicates the shift in the 
boundary conditions. The relevant phase space of two successive systems with $r_i$ and
$r_{i+1}$ is very similar and the ratio of their partition functions,
${\cal Z}(\beta,r_i)/{\cal Z}(\beta,r_{i+1})$, can be efficiently measured 
as the expectation value of the observable $O(U,r_{i+1}) = \exp{({\overline S}(U,r_{i+1})-{\overline S}(U,r_i))}$
on the ensemble of gauge configurations generated with the action 
${\overline S}(U,r_{i+1})$~\cite{DellaMorte:2008jd}. The discrete derivative  
is then written as 
\vspace{-0.625cm}

\be\label{eq:prod}
\frac{1}{2 a}\ln{\frac{Z(L_0,{\bf \xi} + a/L_0 \hat k)}{Z(L_0,{\bf \xi} - a/L_0 \hat k)}}
 = \frac{1}{2 a} \sum_{i=0}^{n-1}
\ln{\frac{{\cal Z}(\beta,r_i)}{{\cal Z}(\beta,r_{i+1})}}\; .
\ee
\vspace{-0.425cm}

\noindent All the details and the results of the computation of 
$Z_T$ will be presented elsewhere \cite{noi:2014big}. In Tab.~\ref{tab:ZT} we report 
the values of $Z_T$ at the 8 values of $\beta$ needed to renormalize the entropy 
density at the temperature $T_0$ computed with shift ${\bf \xi}=(1,0,0)$ and $(1,1,1)$.
Albeit with smaller statistical errors, our values are in agreement with those 
found in Ref.~\cite{Robaina:2013zmb}. Also in this case we have subtracted the discretiazion 
effects of the free theory. In each of the two sets of data we keep $L_0$ fixed 
in physical units, so that residual (small) discretization effects in $Z_T$ will be removed in the 
continuum limit extrapolation of the renormalized entropy density. Discretization effects due 
to finite volume are negligible within our errors. For completeness, in the same Table we also 
report the corresponding expectation values of $\langle T_{0k} \rangle _{\bf \xi}$ in the large 
volume which enters Eq.~(\ref{eq:Snorm}). The results for $s(T_0)/T^3_0$ as defined in 
Eq.~(\ref{eq:Snorm}) are shown in the right plot of Fig.~\ref{fig:cont_extr}.
\begin{table}[t!]
\begin{center}
\begin{tabular}{lcll}
\hline\\[-0.175cm]
$\beta$  & $L_0/a$ &$\;\;\;\;\;\;\;\langle T_{0k} \rangle_{(1,0,0)}$&$\;\;Z_T$\\[0.125cm]
\hline\\[-0.175cm]
$6.0403$ & $3$  & $-5.4278(22)\, 10^{-3}$ & $1.585(6)$\\[0.125cm]
$6.2257$ & $4$  & $-1.7262(5) \, 10^{-3}$ & $1.523(6)$\\[0.125cm]
$6.3875$ & $5$  & $-0.7203(5) \, 10^{-3}$ & $1.497(4)$\\[0.125cm]
$6.5282$ & $6$  & $-0.3536(5) \, 10^{-3}$ & $1.484(6)$\\[0.125cm]
\hline\\[-0.175cm]
\hline\\[-0.175cm]
$\beta$ & $L_0/a$  &$\;\;\;\;\;\;\;\langle T_{0k} \rangle_{(1,1,1)}$&$\;\;Z_T$\\[0.125cm]
\hline\\[-0.175cm]
$6.2670$ & $3$  & $-6.584(11) \, 10^{-4}$ & $1.528(6)$\\[0.125cm]
$6.4822$ & $4$  & $-2.187(3)  \, 10^{-4}$ & $1.475(6)$\\[0.125cm]
$6.6575$ & $5$  & $-0.9251(19)\, 10^{-4}$ & $1.456(3)$\\[0.125cm]
$6.7981$ & $6$  & $-0.4524(14)\, 10^{-4}$ & $1.439(6)$\\[0.125cm]
\hline\\[-0.175cm]
\end{tabular}
\vspace{-0.25cm}

\caption{
\label{tab:ZT} The bare vacuum expectation values of 
$\langle T_{0k} \rangle_{\bf \xi}$ at the reference temperature
$T_0$ for ${\bf\xi}=(1,0,0)$ and $(1,1,1)$. The renormalization 
constant $Z_T$ at the corresponding eight $\beta$ values is also
reported.}
\end{center}
\vspace{-1.0cm}

\end{table}
The typical statistical error is just below half a percent, while the
largest discretization error is roughly $3\%$. The continuum limit
extrapolation 
of the data with ${\bf \xi}=(1,0,0)$ and $(1,1,1)$ 
at the three finer lattices are in excellent agreement among themselves. 
A combined extrapolation gives $s(T_0)/s_{\rm SB}(T_0)=0.788(4)$ 
with a $\chi^2/{\rm dof}=0.74$, see Tab.~\ref{tab:continuum}.\\
\noindent {\it Results and conclusions.---} 
Once the entropy density has been measured at $T_0$,
$s(T)$ at the other temperatures is computed by solving the straightforward recursive relation for the 
step-scaling function. The values obtained for the entropy density are reported in 
Tab.~\ref{tab:continuum} and shown in Fig.~\ref{fig:cont_res}. The precision
reached for $s(T)$ is half a percent at $T_0$, and becomes at most $1.5\%$ at 
$T/T_0=8\sqrt{2}$. We expect to reduce the latter error to the same level of the former 
once the renormalization constant is determined in the full range 
$0\leq g_0^2\leq 1$ \cite{noi:2014big}. Taking into account that the entire computation
required a few million of core hours on BG/Q, the precision reached shows the 
potentiality of the strategy.\\ 
\indent The results for the entropy density are in agreement with those
in Refs.~\cite{Boyd:1996bx,Meyer:2009tq}, and for $T > 2\, T_c$ with the 
more precise ones in Ref.~\cite{Borsanyi:2012ve}. Our data differ by several 
standard deviations from those in Ref.~\cite{Borsanyi:2012ve} in the interval
$T_c < T < 2\, T_c$. A more detailed comparison will be presented in Ref.~\cite{noi:2014big}, 
where more points will be added in this low-temperature region. The step-scaling 
function at $T\sim\! 15 T_c$ is already compatible with the high-temperature limit
within the half a percent uncertainty quoted. The entropy density, however, 
still differs from the Stefan-Boltzmann value by rougly $5\%$ at $T\simeq 20\, T_c$. To compare 
with the known perturbative formula~\cite{Kajantie:2002wa}, we use 
$\Lambda_{\msbar}\, r_0=0.586(48)$~\cite{Capitani:1998mq,Necco:2001xg} and we fix the $O(g^6)$ 
undetermined coefficient by matching the perturbative value of the entropy density 
with our data at the largest temperature $T\simeq 20\, T_c$. The results are shown in 
Fig.~\ref{fig:cont_res}. Despite the good agreement, it must be said that the
contribution from the various orders in the perturbative series is oscillating. At 
our largest temperature the contribution of $O(g^6)$ is roughly $40\%$ of the total correction
to the entropy density given by the other terms, see Ref.~\cite{noi:2014big} for
more details.\\
\indent On a more theoretical side, the results presented in this Letter
are a direct non-perturbative verification of the consequences of Lorentz invariance 
at finite T.\\
\noindent {\it Acknowledgments.---} 
We thank H.~B.~Meyer and D. Robaina for discussions. The simulations were 
performed on the BG/Q at CINECA (INFN and LISA agreement), and on PC clusters at the 
Physics Department of the University of Milano-Bicocca. We thankfully acknowledge the 
computer resources and technical support provided by these institutions. 
This work was partially supported by the INFN SUMA project.
\vspace{-0.625cm}

\bibliography{./lattice}

\end{document}